# Passing Waves from Atomistic to Continuum

Xiang Chen[1], Adrian Diaz[1], Liming Xiong[2], David L. McDowell[3,4] and Youping Chen[1]

[1]Department of Mechanical and Aerospace Engineering, University of Florida, Gainesville, FL 32611
[2]Department of Aerospace Engineering, Iowa State University, Ames, IA 50011
[3]Woodruff School of Mechanical Engineering, Georgia Institute of Technology, Atlanta, GA 30332, USA
[4]School of Materials Science and Engineering, Georgia Institute of Technology, Atlanta, GA 30332, USA

**Abstract**

Progress in the development of coupled atomistic-continuum methods for simulations of critical dynamic material behavior has been hampered by a spurious wave reflection problem at the atomistic-continuum interface. This problem is mainly caused by the difference in material descriptions between the atomistic and continuum models, which results in a mismatch in phonon dispersion relations. In this work, we introduce a new method based on atomistic dynamics of lattice coupled with a concurrent atomistic-continuum method to enable a full phonon representation in the continuum description. This then permits the passage of short-wavelength, high-frequency phonon waves from the atomistic to continuum regions. The benchmark examples presented in this work demonstrate that the new scheme enables the passage of all allowable phonons through the atomistic-continuum interface; it also preserves the wave coherency and energy conservation after phonons transport across multiple atomistic-continuum interfaces. This work is the first step towards developing a concurrent atomistic-continuum simulation tool for non-equilibrium phonon-mediated thermal transport in materials with microstructural complexity.

## 1. Introduction

Over the past two decades, extensive research efforts have been devoted to the development of coupled atomistic-continuum methods. A key challenge in nearly all such concurrent multiscale methods has been to interface atomistic models with continuum mechanics. Across this interface there is a change in material description, governing equations and numerical resolution, leading to "ghost forces" in static multiscale methods, and "spurious wave reflections" in dynamic multiscale simulations. While significant progress has been made in reducing or eliminating the ghost forces [1-3], the spurious wave reflection at the atomistic-continuum (A-C) interface is still the major obstacle to the development of dynamic multiscale methods. As a result, most of the efforts in the development of dynamic concurrent multiscale methods have been devoted to minimizing reflected waves or absorbing heat at the atomistic-continuum interface [4-9].

One early method that addresses the spurious wave reflection problem is to modify the atomistic equation of motion based on Langevin dynamics using a stadium damping to account for the entropy loss in the coarse-grained (CG) region. Based on the generalized Langevin equation (GLE), Cai et al. [5] calculated the time-history-kernel (THK) matrix to minimize the wave reflection in a one-dimensional (1D) model that couples an atomistic simulation system to linear elastic surroundings. Through a wave-packet test, their approach was shown to significantly reduce wave reflections at the A-C interface. This approach was then employed by Wagner and Liu in the bridging scale method (BSM), also for a 1D case [10]. The THK-based method, although effective in eliminating spurious waves, requires intensive computational cost and is difficult to extend to higher dimensions. Therefore, within the framework of BSM, attempts have been made to derive a compact THK that is computationally much less expensive. However, the compact THK is only effective for linear continua and its extension to nonlinear solids remains challenging [6-8,11]. Another two methods in the category of minimizing the wave reflection at the A-C interface are: (1) to treat the A-C interface issue as an optimization problem [4,9], and (2) to apply the idea of digital filters in the signal processing field to selectively filter out the short-wavelength phonon waves reflected back into the atomistic region [12,13].

The above mentioned approaches of eliminating spurious wave reflections essentially assume that minimizing, absorbing, or filtering out the unwanted wave reflections at the A-C interface is sufficient for purpose of multiscale materials modeling [14,15]. However, as pointed out by Chirputkar and Qian [16], any interface treatment that involves damping to eliminate wave reflections will also dissipate the fine-scale wave components. The fine-scale wave information needs to be transmitted, instead of simply being minimized or filtered out, from the atomic into the continuum domain [16]. To address this issue, Qian and his coworkers proposed a concurrent multiscale methodology through integrating an enrichment function into the basic framework of a space-time discontinuous Galerkin finite element (FE) [16]. This method was applied to one-dimensional wave propagation and two-dimensional crack propagation problems [17]. The numerical examples demonstrate the robustness of their method in terms of both energy conservation and the almost-nonreflective interface. However, their formulation requires solution for extra degrees of freedom in order to include short-wavelength wave components in the FE region. In addition, at each time step, the enriched interpolation functions must vanish at the FE nodes. As a result, it is difficult to preserve the correct phasing of waves after transmitting across the A-C interface. Therefore, this method is not readily applicable to dynamic problems for which phonon coherency is important, e.g., the transient phonon thermal transport in materials under an ultrashort laser pulse. The ultrashort laser induced heat pulse has been widely used in pump-probe techniques to characterize the thermal transport properties of materials [21-26]. It has been pointed out that coherence is the key property of a laser pulse, and is a general phenomenon that takes place whenever the ultrashort lasers interact with solid materials [18]. Moreover, in such experiments, the phonon-microstructure scattering involves wavelengths at the mesoscale, which are beyond the reach of fully atomistic simulations to date. This necessitates a multiscale method that not only allows all possible phonons to pass the A-C interface, but also maintains the phonon wave coherency.

This work aims to address the spurious phonon wave reflection problem within the framework of the concurrent atomistic-continuum (CAC) method. CAC is a dynamic multiscale method with a unified formulation of balance laws that govern both atomistic and continuum modeling regions [19-21]. Therefore, it reduces the phonon wave reflection problem at the A-C interface to a numerical problem caused by different FE mesh size. This is a well-known and long-standing problem in the use non-uniform FE meshes and was regarded as one of the major unsolved problems of the Finite Element Method (FEM) by Zienkiewicz in his "Achievements and Some Unsolved Problems of the Finite Element Method" [22]. The objective of this work is to formulate and implement a lattice dynamics (LD)-based FE scheme to demonstrate the ability of CAC to accommodate short-wavelength, high-frequency phonon waves in the coarse-grained (CG) domain through the addition of a supplemental basis for the FE solution. The new scheme will facilitate the passage of a full population of phonon waves from the atomistic to the CG region without introducing any extra degrees of freedom. The remainder of this article is organized as follows: in Section 2, we briefly introduce the CAC method and quantify the phonon wave transmission and reflection in CAC models that employ the conventional linear interpolation; in Section 3, we present the formulation of the LD-based FE scheme, the results of the benchmark examples, and the error analysis. The paper then concludes with a brief summary and discussion in Section 4.

## 2. The CAC method and wave reflections at the A-C interface

The formulation of the CAC method is an extension of Irving-Kirkwood's nonequilibrium statistical mechanical formulation of hydrodynamics equations [23] to a concurrent atomistic-continuum formulation for crystalline materials with a two-level structural description of the materials. In CAC, a crystal is described as a continuous collection of lattice cells, but embedded within each cell is a group of discrete atoms, similar to the two-level material description in Micromorphic Theory [24-27]. The concurrent two-level materials description leads to a multiscale representation of the balance laws [19,20]. Supplemented by the underlying interatomic potential, the reformulated balance laws solve for both the continuous lattice deformation, and the rearrangement of atoms within each lattice cell, naturally leading to a concurrent atomistic-continuum methodology. A chief attribute of the CAC formulation is the single set of governing equations that governs the material behavior in both fine and coarse-scale regions [28]. This distinguishes CAC from the other concurrent multiscale methods that merely abut the atomistic and continuum regions with distinct governing equations. By including the internal degrees of freedom of atoms embedded within each lattice cell in the formulation, CAC can be used to simulate both monatomic and polyatomic crystalline materials [2,29,30]. The CAC method has been demonstrated to be able to predict the dynamics of dislocations [31-35] and fracture [2,36,37], and to reproduce full sets of phonon branches in both atomistic and coarse-grained regions [38] above the wavelength of coarse-grained elements.

The current CAC simulator employs the conventional FE with tri-linear interpolation functions. To quantitatively demonstrate the nature of wave reflections at the A-C interface, we constructed a CAC model which contains both an atomistically resolved region and a coarse-meshed FE region for a 1D atomic chain. Figure 1 presents the schematic of the computer model, which has 500 linear finite elements in the middle representing the coarse-grained (CG) region, and 1500 atoms at each of the two ends representing the atomistic region. Periodic boundary conditions are applied at the two ends of the chain. The element size $h$ in the CG region is set to be $6a$, where $a$ is the lattice constant. The Lenard-Jones potential [39] is used for the interatomic interaction in both the atomistic and CG domains, without loss of generality.

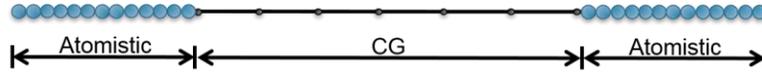

**FIG. 1 The schematic sketch of a CAC model for a 1D atomic chain.**

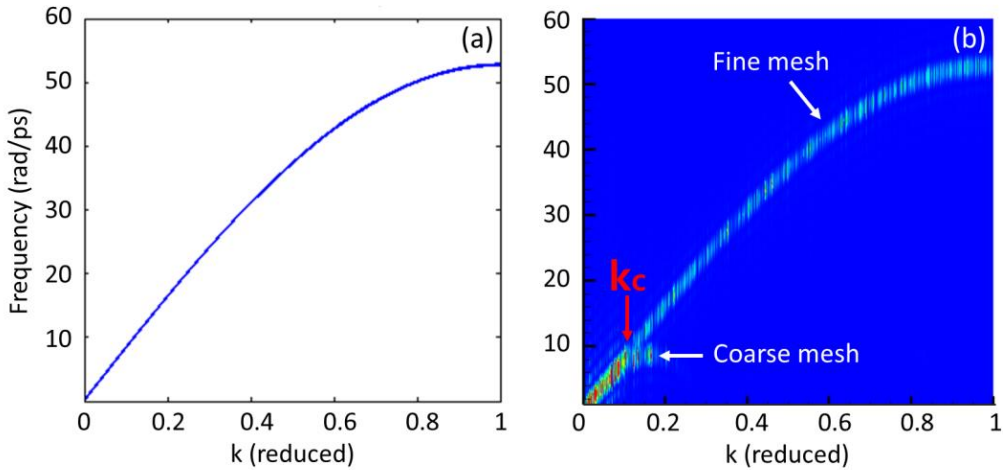

**FIG. 2 (a) Phonon dispersion relation of a monatomic chain calculated from lattice dynamics, and (b) phonon spectrum energy density calculated from the CAC simulation.**

Figure 2a shows the analytical results of the phonon dispersion relation for a monatomic chain from LD calculations. Figure 2b presents the phonon spectrum energy density calculated from the CAC model presented in Fig. 1. Here the phonon spectrum energy density, defined as the average kinetic energy per unit cell as a function of wavevector and frequency [40], is calculated through postprocessing the velocities of atoms in the atomistic region, and the FE nodal velocities in the finite element region. From the phonon spectrum in Fig. 2b, we can clearly identify the phonon dispersion curves for both coarse and fine regions. On one hand, we see the phonon dispersion relation obtained for the atomically resolved regions in the CAC model is identical to that from the LD calculation in Fig. 2a. This demonstrates the CAC method can reproduce the exact dynamics of an atomistic system if the finest FE mesh ($h = a$) is employed. On the other hand, it is seen that the phonon dispersion relation of the CG region only overlaps that of the atomistic region for wavevectors smaller than a critical value. This is because coarse-graining cuts off the small wavelength phonons with wavevectors larger than the critical wavevector $k_c$; $k_c$ is determined by the finite element size $h$ according to Eq. (1) with an allowable error ε, i.e.,

$$k_c = \max_k \left\{ \left| \sin\left(\frac{ka}{2}\right) - \sin\left(\frac{kh}{2}\right) \right| \le \varepsilon \right\}$$
(1)

It is seen from Fig. 2b that the difference between the phonon dispersion relation in the atomistic domain and that in the CG domain is negligible for $k < k_c$. This assures that even with a linear interpolation function a CAC model discretized with coarse-meshed finite elements can precisely predict the dynamics of dispersive phonons whose wavelengths are larger than $2\pi/k_c$.

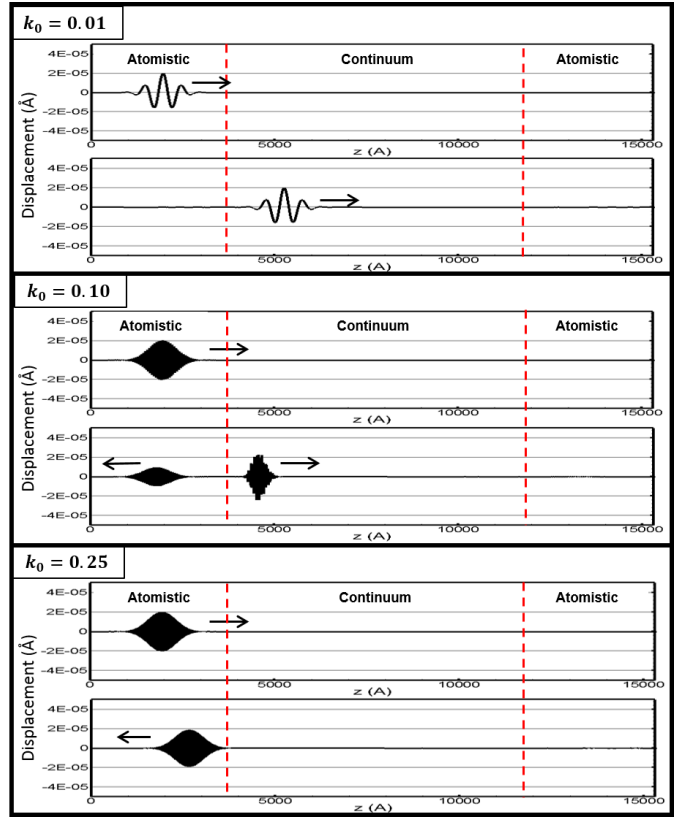

**FIG. 3 CAC simulation of phonon wave packet dynamics in a 1D atomic chain, with (a) $k_o = 0.01$, (b) $k_o = 0.1$, and (c) $k_o = 0.25$. The arrows indicate the propagation direction of the wave-packet, and the red dashed lines mark the location of the A-C interface.**

To quantify the wave transmission and reflection at the A-C interface, we conducted CAC simulations of phonon wave-packet dynamics. Three sets of CAC simulation results are presented in Fig. 3 with wave-packets of three different central wavevectors, $k_o = 0.01, 0.1$ and $0.25$. Essentially, these three wave-packets represent three typical points on the phonon dispersion curve as shown in Fig. 2. $k_o = 0.01$ is within the region where the atomistic and CG dispersion curves overlap, $k_o = 0.1$ is slightly beyond the overlap region, and $k_o = 0.25$ is a short-wavelength, high-frequency phonon wave packet that cannot be modeled by the CG domain using the linear interpolation function. The phonon wave-packets are constructed following the method by Schelling et al., and Aubry et al. [41-43]. The basic idea is to create a localized phonon wave-packet through a superposition of normal modes.

It is seen from Fig. 3 that the wave-packet with a small $k$ (e.g., $k_o = 0.01$) transmits across the A-C interface smoothly. This again demonstrates that CAC is able to precisely simulate the dynamics of dispersive phonon waves when $k < k_c$. In contrast, when the wavevector is slightly beyond the overlap region of the two phonon dispersion curves (e.g., $k_o = 0.1$), the phonon wave-packet is partially reflected by the A-C interface. When the wavevector $k$ is further increased (e.g., $k_o = 0.25$), i.e., the wavelength is decreased to a value that cannot be allowed by the CG domain using the linear FE interpolation function, the entire wave-packet is reflected by the A-C interface.

The results in Fig.3 confirm that wave-reflection at the A-C interface in CAC solely stems from the numerical resolution mismatch between the atomistic and the CG domain. This is attributed to the fact that in a CAC simulation, the nonlocal internal force field-displacement relationship is the only material constitutive relation employed; consequently, the frequency-wavevector relation is dispersive. However, as the usual FE mesh in continuum models cuts off elastic waves with short wavelength [44], so does the FE mesh with the CAC for dynamic simulations. In order to allow short-wavelength waves to pass from an atomically resolved region to a coarse-meshed FE region in CAC, we must reformulate the FE approximation function in the CG domain to allow the full population of phonon waves from the atomic domain to propagate into the CG domain.

## 3. A lattice dynamics-based finite element scheme

### 3.1 The lattice dynamics formulation

The weak form of the CAC governing equation can be expressed as [34,36,45]

$$\int_{\Omega(x)} \left( \rho^\alpha \ddot{u}^\alpha(x) - \mathbf{f}_{\text{int}}^\alpha(x) \right) \Phi_i(x) \mathrm{d}x = 0 \tag{2}$$

which is simplified to only depend on the internal force density for this demonstration, i.e., the effect of temperature and external force field are ignored; here $x$ is the physical space coordinate of the continuously distributed lattice, $\Omega(x)$ is the domain of $x$. $\alpha$ represents the $\alpha$-th atom in the unit cell located at $x$. $\rho^\alpha$ is the local density of mass, $\mathbf{f}_{\text{int}}^\alpha$ is the internal force density, which is a nonlinear nonlocal function obtained from the interatomic potential function. In Eq. (2), the basis $\Phi_i(x)$ of the unknown displacement field $u^\alpha$ is open to selection based on the problem to be numerically solved. The freedom to choose the basis $\Phi_i(x)$ of the solution allows for the solution of a wider range of problems.

To seek a solution to the spurious wave reflection problem at the A-C interface, in this work, we employed the LD description of traveling waves to formulate a new FE approximation, i.e., the phonon wave-based FE interpolation scheme, to enable the propagation of short-wavelength waves that are cut off by the linear finite element. For this purpose, we suggest the union of a FE basis and a modal superposition limited to a range of discrete frequencies as the solution of the global displacement field in the coarse-grained domain. The conventional FE approximation functions allow the description of waves with a wavelength longer than the critical wavelength. The LD formulation is to describe the waves with a wavelength shorter than the critical wavelength. Considering a general polyatomic crystalline system with $N^\alpha$ atoms in the unit cell, the atomic-scale displacements can be approximated as

$$u^\alpha(X,t) = \sum_{i=1}^{2^d} N_i(X)[U_i^\alpha(t) - U_{si}^\alpha(t)] + u_s^\alpha(X,t), \quad \alpha = 1,2,\ldots,N^\alpha \tag{3}$$

in which, $u_s^\alpha(X,t)$ is the combination of short-wavelength phonons with $k > k_c$ (denoted by the subscription $s$); $N_i(X)$ is the conventional FE tri-linear shape function, $d$ is the dimensionality of the computer model, $2^d$ is the total number of nodes in an element; $U_i^\alpha(t)$ is the total displacement of the α-th atom at the i-th FE node, and $U_{si}^\alpha(t)$ is the contributions of short wavelength components to the displacements of the α-th atom embedded in the i-th node. The selection of this basis is for demonstrative purposes and it is not meant to be an efficient implementation to enrich the basis of the solution. There is no loss in generality when representing the current displacements at an arbitrary timestep with waves of unknown amplitude. The loss in generality would only come from attempting to model future displacements with the wave solution corresponding to any dispersion characteristics of each of those waves.

It should be noted that the conventional FE method based on a local polynomial approximation is prohibitive in treating problems that have solutions with oscillatory character. For example, in a system described by finite elements with linear approximation, the global relative error scales as $O(k(kh)^2)$, where $h$ is the element size [46]. Hence, when the wavenumber $k$ is increased, reducing $h$ at the same rate is not sufficient to keep the error constant. A higher order polynomial has also been proven to be ineffective when $k$ is large [46]. Thus, a natural way is to incorporate the wave natures of the crystalline dynamics at atomic scale into the approximation subspace of the finite elements.

The motion of an atom in an atomistic system is an oscillatory function of time. Nevertheless, the motion of atoms in crystalline materials turns out to be most readily described not in terms of the individual atoms, but in terms of traveling waves, which are named lattice vibrations by Born [47]. In a lattice dynamics (LD) system with a harmonic approximation, the displacement of atoms can be decomposed into a linear combination of normal modes [48], permitting only a discrete set of wavevectors. The number of allowed wavevectors is equal to the number of primitive unit cells. Considering the contributions from only short-wavelength phonon components (denoted by the subscription $s$), i.e., $k > k_c$, the displacement of the $\alpha$-th atom in the unit cell located at $X$ in the undeformed configuration can be described as

$$\boldsymbol{u}_s^\alpha(\boldsymbol{X}, t) = \frac{1}{(N_l m^\alpha)^{1/2}} \sum_{k,\nu(k>k_c)} \boldsymbol{e}_{k\nu}^\alpha Q_{k\nu} \exp(i(\boldsymbol{k} \cdot \boldsymbol{X} - \omega_{k\nu} t)) \tag{4}$$

Equation (4) decomposes the atomic displacement into a linear combination of normal modes, each representing the contribution from a wave characterized by the phonon branch $\nu$ and the wavevector $\boldsymbol{k}$. $Q_{k\nu}$ is the normal mode coordinate, which has subsumed the time dependence and hence gives both the amplitude of the wave and the time dependence. $\boldsymbol{e}_{k\nu}^\alpha$ is the eigenvector or the polarization vector that determines the direction in which each atom moves, $\omega_{k\nu}$ is the corresponding frequency, $m^\alpha$ is the mass of the $\alpha$-th atom in the $l$-th unit cell, and $N_l$ is the total number of unit cell in the system. Eq. (4) can be rewritten as

$$\boldsymbol{u}_s^\alpha(\boldsymbol{X}, t) = \sum_{k,\nu(k>k_c)} U_{k\nu}^\alpha \boldsymbol{e}_{k\nu}^\alpha \exp(i(\boldsymbol{k} \cdot \boldsymbol{X} - \omega_{k\nu} t)), \tag{5}$$

where $U_{k\nu}^\alpha = \frac{1}{(N_l m^\alpha)^{1/2}} Q_{k\nu}$.

Therefore, $\boldsymbol{U}_{si}^\alpha(t)$ in Eq. (3) can be expressed as

$$\boldsymbol{U}_{si}^\alpha(t) = \sum_{k,\nu(k>k_c)} U_{k\nu}^\alpha \boldsymbol{e}_{k\nu}^\alpha \exp(i(\boldsymbol{k} \cdot \boldsymbol{X}_i - \omega_{k\nu} t)). \tag{6}$$

Since the conventional tri-linear interpolation functions satisfy the "partition of unity" relation, i.e., $\sum_{i=1}^{2^d} N_i(\boldsymbol{X}) = 1$, Eq. (3) can be re-written as

$$\boldsymbol{u}^\alpha(\boldsymbol{X}, t) = \sum_{i=1}^{2^d} N_i(\boldsymbol{X})[\boldsymbol{U}_i^\alpha(t) - \boldsymbol{U}_{si}^\alpha(t) + \boldsymbol{u}_s^\alpha(\boldsymbol{X}, t)] \tag{7}$$

or in a more condensed form,

$$\boldsymbol{u}^\alpha(\boldsymbol{X}, t) = N_i(\boldsymbol{X}) \boldsymbol{V}_i(\boldsymbol{X}, t), i = 1, \dots, 2^d, \tag{8}$$

in which, the summation over repeated indices is implied, and

$$\boldsymbol{V}_i(\boldsymbol{X}, t) = \boldsymbol{U}_i^\alpha(t) - \boldsymbol{U}_{si}^\alpha(t) + \boldsymbol{u}_s^\alpha(\boldsymbol{X}, t). \tag{9}$$

In Eq. (9), the crucial components to be determined are $\boldsymbol{u}_s^\alpha(\boldsymbol{X}, t)$ and $\boldsymbol{U}_{si}^\alpha(t)$, which require the calculation of the amplitude $U_{k\nu}^\alpha$ of each short-wavelength phonon mode.

## 3.2 Calculation of the amplitudes of short-wavelength phonon modes by FFT

The objective of this work is to modify the conventional displacement approximation in the FE region by including the short-wavelength phonon waves, the amplitude of which, denoted as $U_{k\nu}^\alpha$, can be obtained from the analysis of the displacements in the atomistic region within a CAC model. In order to solve for the amplitudes, one must take the 3D

Fourier transform of the system one atom in the unit cell at a time. Specifically, we wish to represent the set of displacements in the following discrete atomic description:

$$\boldsymbol{u}_s^\alpha(\boldsymbol{r}_i, t) = \sum_{k,\nu} A_{k\nu} \boldsymbol{e}_{k\nu}^\alpha \exp(\boldsymbol{k} \cdot \boldsymbol{r}_i - \omega_{k\nu} t) + B_{k\nu} \boldsymbol{e}_{k\nu}^\alpha \exp(\boldsymbol{k} \cdot \boldsymbol{r}_i + \omega_{k\nu} t), \qquad (10)$$

where $\boldsymbol{r}_i$ is the equilibrium lattice position for which the displacement is being calculated. $A_{k\nu}$ and $B_{k\nu}$ are the unknown coefficients to be computed for each mode. In the atomistic region of interest, we have the initial positions, displacements $\boldsymbol{u}_s^\alpha(\boldsymbol{r}_i, t)$, and initial velocities $\boldsymbol{v}_s^\alpha(\boldsymbol{r}_i, t)$. In order to calculate the amplitudes, we must exploit the periodic arrangement of the lattice sites one atom in the unit cell each time by using the fast Fourier transform (FFT). This will in turn give us the amplitude vector $\boldsymbol{C}_k^\alpha$ of the mode $\boldsymbol{k}$ for the set of the $\alpha$-th atom in every lattice site. The FFT is taken as follows:

$$\boldsymbol{C}_k^\alpha = \sum_j \boldsymbol{u}_j^\alpha \exp(-\boldsymbol{k} \cdot \boldsymbol{r}_j), \qquad (11)$$

where the index $j$ is used to sum the displacement over all unit cells for the $\alpha$-th atom in order to perform the FFT. The phonon modes in Eq. (10) evaluated at $t = 0$ for a specific wave vector allow us to relate the modal amplitude computed in (11) to the modal amplitude for a specified atom within the unit cell at a specific wave vector, i.e.,

$$\boldsymbol{C}_k^\alpha = \sum_\nu (A_{k\nu} + B_{k\nu}) \boldsymbol{e}_{k\nu}^\alpha. \qquad (12)$$

This gives us half of the required equations to solve the initial value problem. The rest will require the FFT of the velocities to complete the initial value problem. The amplitudes for the modal description of velocity can then be equated to the amplitudes present in the time derivative of Eq. (12) at time zero, i.e.,

$$\boldsymbol{D}_k^\alpha = \sum_\nu (-i\omega_{k\nu} A_{k\nu} \boldsymbol{e}_{k\nu}^\alpha + i\omega_{k\nu} B_{k\nu} \boldsymbol{e}_{k\nu}^\alpha). \qquad (13)$$

where $\boldsymbol{D}_k^\alpha$ represents the FFT amplitude for the velocity data. Every wavevector corresponds to $2\nu$ unique coefficients ($A_{k\nu}$ and $B_{k\nu}$). The Fourier transform of position and velocity for a certain wavevector $\boldsymbol{k}$, and for atoms $\alpha$ in all unit cells provides us with a $2\nu$ by $2\nu$ system for the coefficients $A_{k\nu}$ and $B_{k\nu}$ that correspond to each wavevector.

### 3.3 Benchmark Examples

To test the formulation and the numerical implementation of the LD-based FE scheme, we compare the transmission in the wave-packet simulation based on the conventional linear interpolation implementation of CAC and that with the new interpolation scheme. The 1D wave-packet test shown in Fig. 3 with $k_o = 0.25$ is taken as a numerical example. For the sake of the demonstration, the interatomic force field is simplified to its linear term and only the nearest neighbor interaction is considered.

In Fig. 4a we plot the time sequences of the wave-packet propagation obtained from a CAC simulation using the linear interpolation function (left column), and compare it with a corresponding reciprocal domain analysis in Fig. 4b (right column). A wavelet method [49] is implemented for the reciprocal domain analysis and provides the spatial and temporal information as well as the instant wavevector associated with the wave-packet propagation. Results in Fig. 4 show that with the conventional linear interpolation function the wave-packet is fully reflected by the A-C interface with zero transmission.

In Fig. 5 we present the simulation results obtained from CAC furnished with the new LD-based FE scheme. It is seen that the new interpolation scheme allows the entire phonon wave-packet to transmit across the first A-C interface without any noticeable reflection (less than 0.5%), which is in a sharp distinction from the zero transmission in the simulation with linear interpolation as shown in Fig. 4. Furthermore, the new FE scheme not only enables the A-C interface to pass the short-wavelength phonon from the first atomistic region to the continuum region, but also allows the continuum region to carry the short-wavelength phonons and pass the full information into the next atomistic region with 100% transmission. The complete reflection presented in Fig. 4, and the complete transmission in Fig. 5 provide a clear-cut demonstration and validation of the formulation and numerical implementation.

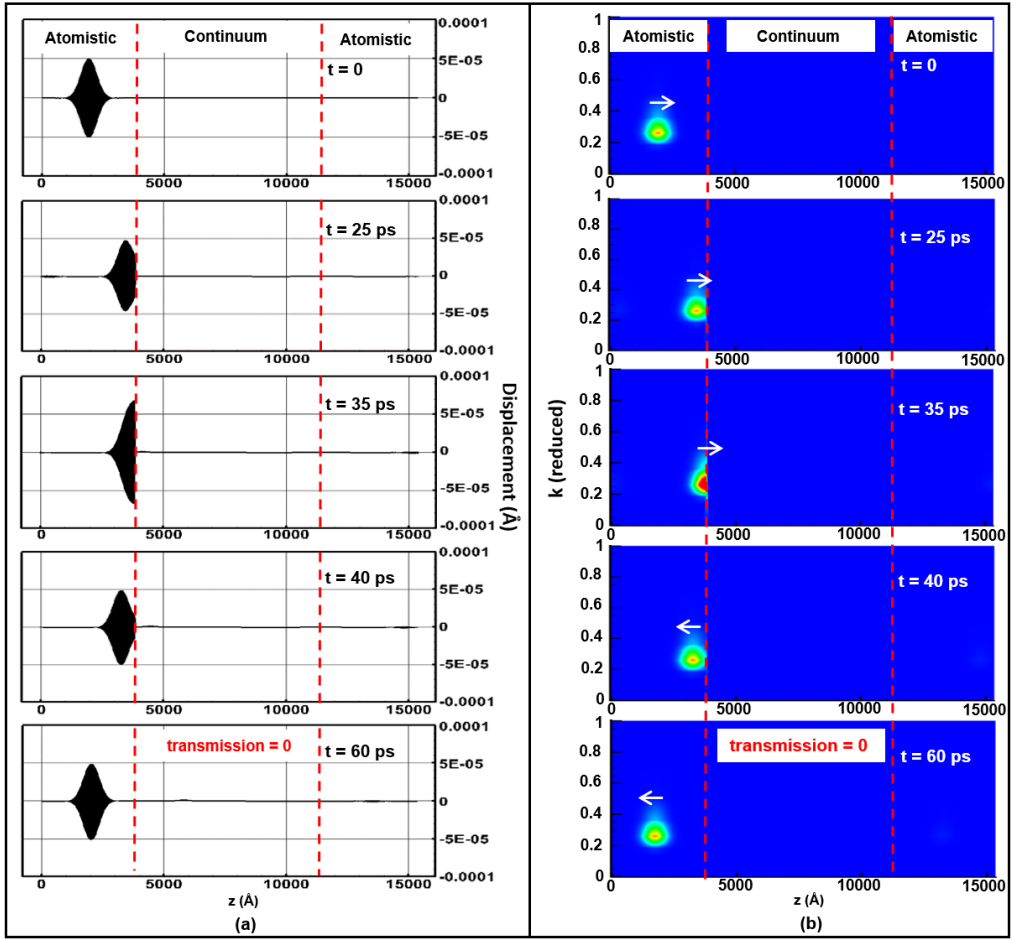

**FIG. 4 Wave-packet ($k_o = 0.25$) reflection by the A-C interface in a CAC simulation with a linear interpolation, (a) the time sequence of the spatial distribution of the displacement (b) the time sequence of the spatial distribution of the wavevector $k$ from a wavelet analysis.**

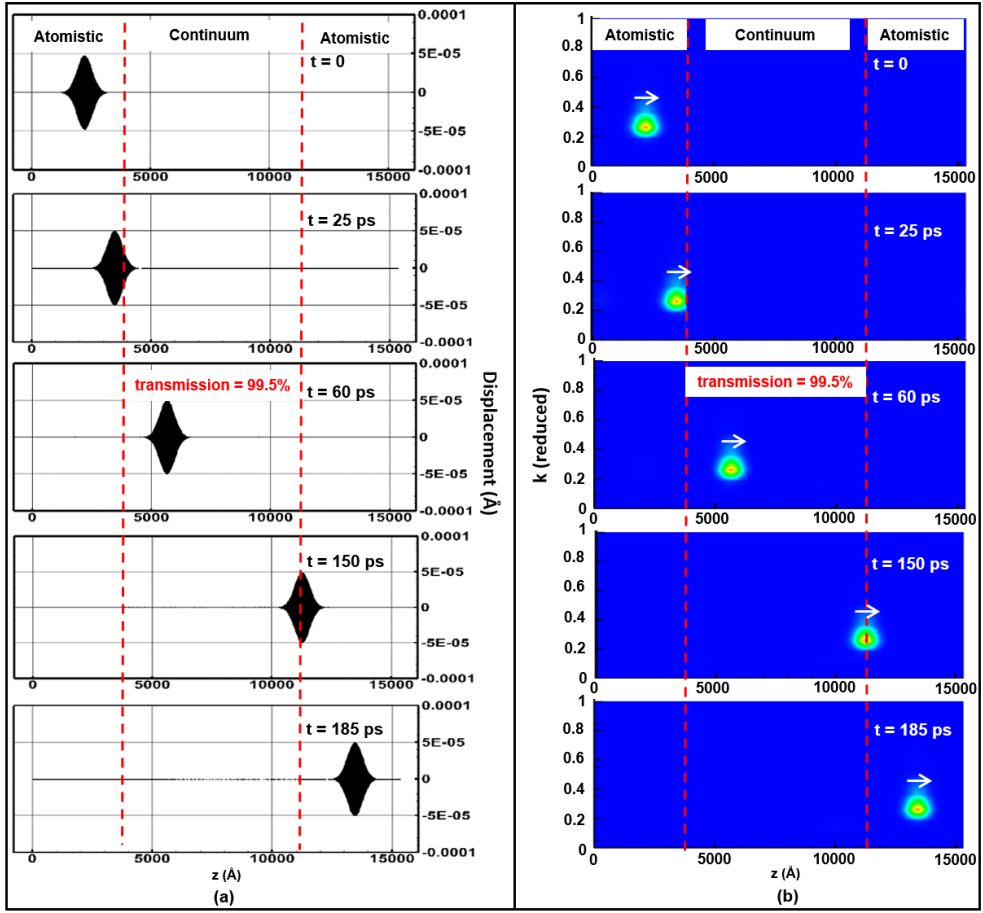

**FIG. 5** Wave-packet ($k_o = 0.25$) transmission across the A-C interface in a CAC simulation with the lattice dynamics-based FE scheme, (a) the time sequence of spatial distribution of the displacement (b) the time sequence of the spatial distribution of the wavevector $k$ from a wavelet analysis.

### 3.4 The source of the error

In the previous section, the LD-based FE scheme has been shown to allow the wave-packet with $k_o = 0.25$ to transmit across the A-C interface without any discernable reflection. Only after changing the threshold of the contour by two orders of magnitude are we able to identify a negligible reflection, which is less than 0.5%. In addition, the wavevectors associated with both the transmitted wave and the reflected wave remain the same as the original wave packet, i.e., $k = 0.25$. By investigating the dependence of the transmission on time-step, the source of this error is identified to be the numerical instability associated with linear system, as proven by the eigenvalue criterion [50,51]. Taking the wave-packet with $k_0 = 0.25$ as an example again, we set up simulations with different time-steps, and the results are presented in Table 1. It is seen that the growth of error can be reduced by setting the time-step smaller (eigenvalues of the update matrix converge closer to a magnitude of one).

**TABLE 1** Effect of the time-step on the phonon wave transmission

| Time-step (fs) | 5.0 | 2.5 | 1.0 | 0.5 |
|---|---|---|---|---|
| Energy transmission (%) | 81.0 | 90.0 | 95.2 | 99.5 |

CAC simulations of the propagation of phonon wave-packets with even larger wavevectors, e.g., shorter wavelength and higher frequency, were also performed in this work. Results show the transmission decreases with an increasing wavevector (i.e., decreasing wave length). For example, the transmission is found to be 97.1% for the wave-packet with $k_o = 0.5$ and 86.2% for $k_o = 0.8$. The loss of transmission with the increase of the wavevector is mainly due to the significantly increased number of time steps required for the large-wavevector phonon waves to pass the A-C interface. This comes as no surprise since phonons with larger wavevectors have lower group velocities; hence, the numerical error has more time to build up before the measurement of the transmission coefficient is taken. The inclusion of anharmonic LD will constitute a future study containing higher order terms in the interatomic force field, in which case the numerical instability can be removed.

## 4. Summary and Discussion

In summary, we have demonstrated the nature of the wave reflection problem in the concurrent atomistic-continuum (CAC) method using a conventional finite element (FE) approximation. The phonon spectrum energy density calculation shows that there is a critical wavevector $k_c$ below which the phonon dispersion curve of the coarse-meshed region overlaps that of the atomically resolved region which is determined by the size of FE in the continuum region. The dynamic behavior of the phonon waves passing through the atomistic-continuum (A-C) interface has been quantified through wave-packet simulations. The simulation results show the long wavelength phonons whose wavevectors are smaller than $k_c$ can transmit across the A-C interface transparently. The medium wavelength phonons whose wavevectors slightly beyond $k_c$ can partially transmit across the A-C interface. The small wavelength phonons whose wavevectors are not allowed in the coarsely-meshed FE domain undergo zero transmission. The results obtained from the calculation of phonon dispersion relation, and the simulations of phonon wave-packet dynamics demonstrate that: (1) the CAC method with the conventional finite element interpolation function is able to predict the correct dynamics of long-wavelength waves, and (2) the wave-reflection at the A-C interface in CAC solely stems from the numerical resolution mismatch between the atomistic and the CG domains.

We then formulate a lattice dynamics (LD)-based FE scheme in order for the CAC method to allow the passage of short-wavelength and high-frequency phonon waves from the atomistic to continuum domains. This formulation is the first attempt towards passing a full population of phonon waves from the atomistic to continuum domain without introducing extra degrees of freedom. A wave-packet simulation demonstrates that with the new FE scheme even the short-wavelength phonon with $k_o = 0.25$ can be transmitted across the A-C interface without a noticeable reflection. Furthermore, the new method preserves the complete phonon information after propagating across multiple atomistic-continuum and continuum-atomistic interfaces. This is an important step forward for the application of the concurrent multiscale method to simulations of phonon thermal transport in materials with microstructural complexity.

It should be noted that in this work the interatomic force field is simplified, for the sake of the demonstration, to its linear term, and we consider only the nearest neighbor interaction. Higher order terms and long-range interactions can be readily implemented in CAC, but they would complicate the demonstration. Adding higher order terms would introduce some anharmonic error with respect to the atomic model that may obscure any other errors in the implementation. Considering larger neighbor lists would require a more complicated LD solution, and thus the simplest neighbor list was chosen in this work.

We would also like to mention that the goal of this work is not to find an efficient implementation for a complex problem; rather, it represents a first attempt that demonstrates the compatibility of the new finite element scheme and the CAC method, which is a starting point towards establishing a tool for dynamic phonon-mediated thermal transport. A future study will extend the numerical implementation to three-dimensional crystalline materials. This will serve to investigate the phonon dynamics in crystalline materials with microstructures, i.e., polycrystals, superlattices, phononic materials, materials with dislocations, etc. Further development based on the introduction of subspaces of the other allowable high-frequency phonon modes will serve to improve the practicality of the approach. The resulting formalism will also possess the potential to broadly improve any concurrent multiscale simulation of thermal transport.

**Acknowledgements**

This material is based upon research supported by the U.S. Department of Energy, Office of Basic Energy Sciences, Division of Materials Sciences and Engineering under Award # DE-SC0006539.